\documentclass{Interspeech}
\usepackage{multirow}
\usepackage[table]{xcolor}
\usepackage{xcolor} 
\usepackage{amsfonts} 
\usepackage{amssymb}
\usepackage{pifont}
\newcommand{\cmark}{\ding{51}}%
\newcommand{\xmark}{\ding{55}}%

\interspeechcameraready

\title{DiEmo-TTS: Disentangled Emotion Representations via Self-Supervised Distillation for Cross-Speaker Emotion Transfer in Text-to-Speech}

\author[]{Deok-Hyeon}{Cho}
\author[]{Hyung-Seok}{Oh}
\author[]{Seung-Bin}{Kim}
\author[affiliation={{\dagger}}]{Seong-Whan}{Lee}

\affiliation[nocounter]{Department of Artificial Intelligence}{Korea University}{Seoul, Korea}
\email{dh\_cho@korea.ac.kr, hs\_oh@korea.ac.kr, sb-kim@korea.ac.kr, sw.lee@korea.ac.kr \thanks{$^\dagger$Corresponding author}}
\keywords{Text-to-speech, cross-speaker emotion transfer}

\usepackage{comment}

\begin{document}

\maketitle

\begin{abstract}
Cross-speaker emotion transfer in speech synthesis relies on extracting speaker-independent emotion embeddings for accurate emotion modeling without retaining speaker traits. However, existing timbre compression methods fail to fully separate speaker and emotion characteristics, causing speaker leakage and degraded synthesis quality. To address this, we propose DiEmo-TTS, a self-supervised distillation method to minimize emotional information loss and preserve speaker identity. We introduce cluster-driven sampling and information perturbation to preserve emotion while removing irrelevant factors. To facilitate this process, we propose an emotion clustering and matching approach using emotional attribute prediction and speaker embeddings, enabling generalization to unlabeled data. Additionally, we designed a dual conditioning transformer to integrate style features better. Experimental results confirm the effectiveness of our method in learning speaker-irrelevant emotion embeddings.
\end{abstract}

\section{Introduction}
Emotional speech synthesis has advanced significantly in recent years and can be categorized into two paradigms: the same-speaker \cite{li18j_interspeech} and the cross-speaker scenario \cite{jo2023cross, zhang2023iemotts, li2022cross1}. The same-speaker scenario refers to when the speaker identity of the training data matches the synthesized speech, requiring sufficient recordings of the target speaker’s emotions. In contrast, the cross-speaker scenario transfers emotion from the reference to a different target speaker. Advances in deep learning \cite{kim2015abstract, lee2018deep, lee2020continuous} allow emotion synthesis without target recordings, highlighting the role of cross-speaker scenarios.

Reference-based style transfer is commonly used in cross-speaker emotion synthesis, but it requires disentangling speaker information from emotion embeddings to prevent residual speaker traits from affecting the target timbre. However, certain prosodic features are inherently associated with the speaker's identity, making complete disentanglement challenging. Therefore, an effective disentangling method is crucial to adequately separate speaker identity from emotion-related prosodic features, ensuring clear and accurate emotional transfer without compromising the target speaker's timbre. Researchers typically implement the disentanglement method via explicit speaker labels, such as the gradient reversal layer (GRL) \cite{ganin2016domain}, as employed in \cite{jo2023cross, zhang2023iemotts, zaidi22b_interspeech, oh2024durflex}. However, using explicit labels for disentanglement introduces a trade-off between preserving emotional information and separating speaker identity, making hyperparameter optimization challenging and leading to synthesis quality \cite{li2022cross1, zaidi22b_interspeech}. Vector quantization (VQ) \cite{van2017neural} offers an alternative approach to separating information without relying on explicit labels. Although VQ is effective in separating information without explicit labels \cite{zhang2023iemotts, huang2022generspeech}, it often results in unintended information loss and requires complex optimization to balance compression with reconstruction quality. To address these issues, \cite{li2022cross1} proposed an orthogonal loss to compensate for the embedding of emotion for the loss of emotional information caused by the disentanglement of speaker information \cite{ranasinghe2021orthogonal}. Although this approach disentangles speaker information, it relies on multiple complex loss functions, making the optimization process overly complex. Furthermore, due to relying on GRL for disentanglement, its effectiveness is inherently limited between feature separation and synthesis quality,  limiting its ability to fully preserve both speaker identity and emotional expressiveness in the generated speech.

Model conditioning strategies remain an active research area, as human speech naturally varies in timbre and prosody depending on the content being expressed. Effective conditioning is crucial for synthesizing nature, as it ensures that key style characteristics are accurately captured and reproduced. Current approaches often condition acoustic models on prosody using simple operations \cite{cho24_interspeech}. 
However, these methods can obscure the style and make it difficult to model complex variations.
Consequently, synthesized speech is sometimes natural and expressive but often inconsistent in style and quality. This highlights the need for more advanced conditioning methods to better reproduce speaker-specific pronunciation, intonation, and emotional nuances.

To overcome these limitations, we propose DiEmo-TTS, a novel approach for cross-speaker emotion disentangling and transfer. Building on the success of self-supervised learning in speech synthesis \cite{pankov2023dino, hwang2024textless}, our method employs teacher-student emotion encoders trained using a distillation with no labels (DINO) strategy \cite{caron2021emerging}. To minimize emotional information loss during disentanglement and enhance emotional expressiveness in synthetic speech, we introduce cluster-driven sampling for DINO. Additionally, style-adaptive condition modules enhance adaptability to emotions and speaker characteristics. Extensive experiments demonstrate the effectiveness of our method in improving emotional expressiveness while preserving the target speaker's identity. Our audio samples are publicly available\footnote{{https://choddeok.github.io/DiEmo-TTS-Demo/}}. 

\begin{figure*}[!t] 
    \centering
\includegraphics[width=1.0\linewidth]{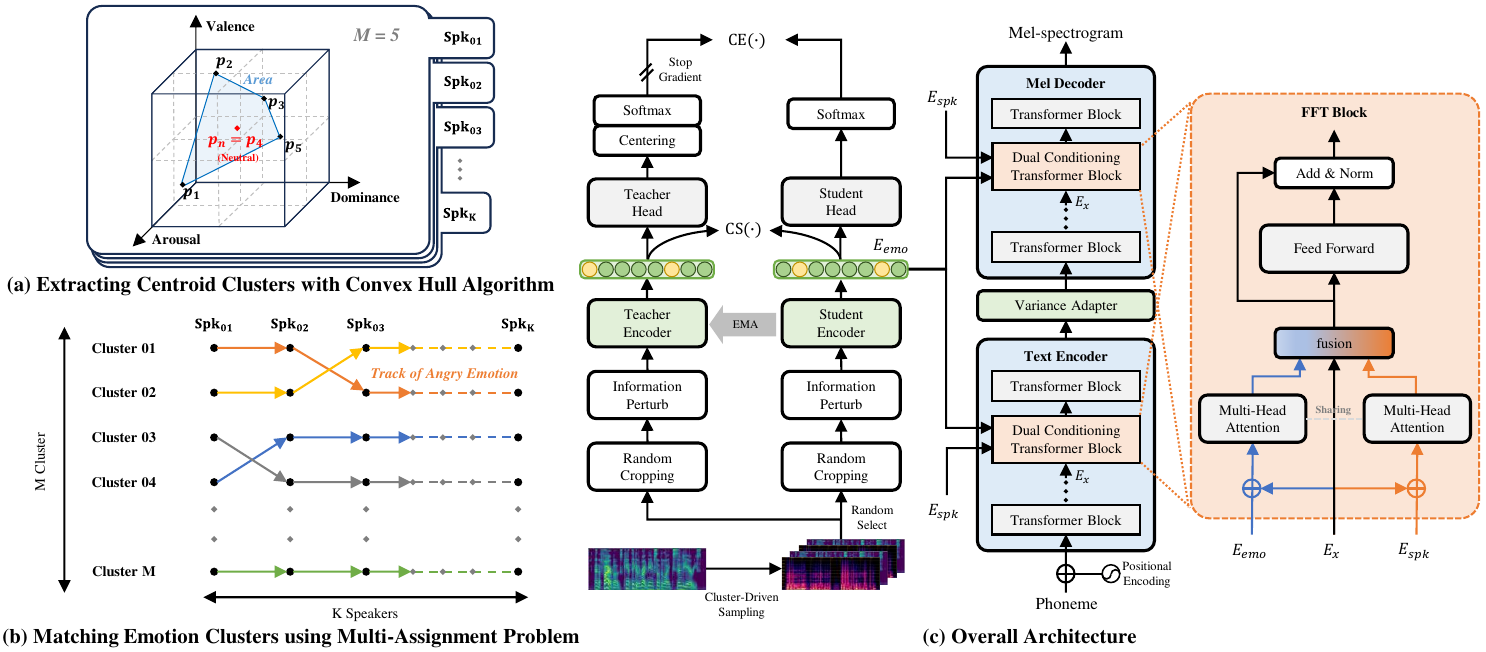}\vspace{-0.3cm}
\caption{Overall framework of DiEmo-TTS.}
\label{model}\vspace{-0.2cm}
\end{figure*}

\section{DiEmo-TTS}
DiEmo-TTS is a cross-speaker emotion transfer system, which improves emotional expressiveness by extracting emotional embeddings via self-supervised distillation. The implementation details are described in the following subsections.

\subsection{Disentanglement of emotion representation via self-supervised distillation }

\subsubsection{Clustering and matching of emotion}
According to \cite{ulgen2024revealing}, speaker embeddings tend to cluster based on different emotional states within the embedding space of each speaker. However, the mapping between emotion categories and intra-speaker clusters is limited due to variations in the speech signal \cite{ulgen2024revealing}. To address this limitation, we introduce emotion dimension pseudo-labels using emotional attribute prediction \cite{goncalves2024odyssey}. In psychology, the emotional state scale positions neutral emotion at the center, while other categorical emotions maintain consistent relative positions within the emotional attribute space \cite{russell1977evidence, zhou2024emotional}. To define a region centered on the neutral emotion cluster $p_{n}$, as shown in Figure \ref{model} (a), we apply the Convex Hull algorithm \cite{barber1996quickhull} to the central points of the predicted emotional attributes for each cluster.
\begin{equation}
p_{n}= \arg\max_{p_{i}\in S }{Area \left ( ConvexHull\left ( S \setminus \{ p_{i} \} \right ) \right )},
\end{equation}
where $S=\{ p_{1},p_{2},...,p_{M} \}$ is the set of mean predicted emotional dimensions, $p_{i}$ represents the mean values of the $i$-th cluster, and $M$ is the total number of clusters. The notation $S \setminus \{ p_{i} \}$ denotes the set difference, meaning the set $S$ excluding the element $\{ p_{i} \}$. Subsequently, a spherical coordinate transformation \cite{cho24_interspeech, 10965917} is performed centered on the neutral emotion cluster $p_{n}$ to extract azimuth and elevation angles, representing the relative positional information of each cluster. Finally, using the relative positional information of azimuth and elevation angles, we align clusters within each speaker to the same emotion. This is achieved through the multiple Hungarian methods for the $k$-assignment problem \cite{gabrovvsek2020multiple}, as illustrated in Figure \ref{model} (b).

\subsubsection{Emotion disentanglement DINO}
Unlike the traditional DINO method, we propose emotion disentanglement DINO, incorporating three approaches to achieve effective emotion disentanglement, as shown in Figure \ref{model} (c). First, emotions within the same cluster are assumed to share the same identity, and cluster-driven sampling is performed accordingly \cite{han2023self}. This cluster-based random cropping enhances data diversity by introducing variations in both content and speakers, even within the same emotion. Instead of emphasizing speaker characteristics as in traditional DINO \cite{pankov2023dino, hwang2024textless, han2023self}, our approach focuses on emotional information while excluding unrelated factors. Second, we apply information perturbation (IP) \cite{zhu2024metts} to distort speaker information while preserving emotional expression. This approach utilizes the correlation between speaker timbre and speech formant, ensuring that formant perturbation effectively distorts speaker identity while preserving emotional expression. Finally, a cosine similarity loss is added to the existing DINO loss to ensure that the emotion embeddings are encoded in a more suitable space. All crops $N$ pass through the student encoder, while only the long crops $L$ pass through the teacher encoder to extract the expressivity embeddings $e_s$, $e_t$. By maximizing the cosine similarity among the embeddings, this approach reduces emotion loss and produces a high-quality embedding space that is more suitable for clustering techniques. Then, the traditional DINO loss is computed as a combination of cross-entropy losses between the head outputs $h_s$ and $h_t$, which are obtained by applying the softmax function. The final formulation $L_{\text{DINO}}$ is structured as follows:
\begin{equation}
\scriptsize
L_{\text{DINO}} = \frac{1}{L \cdot (N - 1)} 
\sum_{l=1}^{L} \sum_{\substack{m=1 \\ m \neq l}}^{N} \left[\text{CE}\left(h_t^l, h_s^m\right) + \text{CS}\left(e_t^l, e_s^m\right) \right],
\tag*{\normalsize (2)}
\end{equation}
where $l$ and $m$ denote the $l$-th long crop and $m$-th all crop, while $s$ and $t$ represent the student and teacher embeddings. Here, $\text{CE}\left (a,b  \right )=-a\text{log}b$ and $\text{CS}\left ( a,b \right )=1-\text{cos}(a,b)$ represent the cross-entropy loss and cosine similarity loss, respectively.

\begin{table*}[!ht]
    \centering 
        \caption{Comparison of our proposed method in cross-speaker emotion transfer.}
    \label{Table1}\vspace{-0.3cm}
        \resizebox{1.00\textwidth}{!}{
    \begin{tabular}{l|cccc|cccc|cccc}
        \toprule
        \multirow{2}{*}{\textbf{Method}} & \multicolumn{4}{c|}{\textbf{Speech Naturalness MOS}} & \multicolumn{4}{c|}{\textbf{Speaker Similarity MOS}} & \multicolumn{4}{c}{\textbf{Emotion Similarity MOS}} \\ 
        \cmidrule{2-13} 
         & \textbf{Angry} & \textbf{Happy} & \textbf{Sad} & \textbf{Surprise} & \textbf{Angry} & \textbf{Happy} & \textbf{Sad} & \textbf{Surprise} & \textbf{Angry} & \textbf{Happy} & \textbf{Sad} & \textbf{Surprise} \\ 
        \midrule
            Expressive FS2 & 3.74$\pm$0.05 & 3.62$\pm$0.05 & 3.74$\pm$0.04 & 3.71$\pm$0.05 & 3.68$\pm$0.05 & 3.69$\pm$0.04 & 3.69$\pm$0.05 & 3.62$\pm$0.04 & 4.10$\pm$0.05 & 3.68$\pm$0.04 & 3.92$\pm$0.05 & 3.98$\pm$0.05 \\ 
        \midrule
        \midrule
            Trans-GRL \cite{jo2023cross} & \textbf{3.77$\pm$0.04} & 3.72$\pm$0.05 & 3.68$\pm$0.04 & 3.70$\pm$0.04 & 3.67$\pm$0.05 & 3.65$\pm$0.04 & \textbf{3.78$\pm$0.05} & 3.67$\pm$0.04 & 3.84$\pm$0.04 & 3.76$\pm$0.05 & 3.76$\pm$0.05 & 3.86$\pm$0.03 \\ 
            Trans-VQ \cite{zhang2023iemotts} & 3.74$\pm$0.05 & 3.77$\pm$0.03 & \textbf{3.71$\pm$0.04} & 3.75$\pm$0.05 & \textbf{3.69$\pm$0.04} & 3.70$\pm$0.04 & 3.74$\pm$0.04 & 3.62$\pm$0.05 & 3.86$\pm$0.05 & 3.71$\pm$0.05 & 3.58$\pm$0.04 & 3.63$\pm$0.05 \\ 
            Trans-Ort \cite{li2022cross1} & 3.67$\pm$0.05 & 3.73$\pm$0.04 & 3.68$\pm$0.04 & 3.79$\pm$0.04 & 3.68$\pm$0.03 & 3.69$\pm$0.04 & 3.73$\pm$0.05 & 3.67$\pm$0.05 & \textbf{3.95$\pm$0.05} & 3.67$\pm$0.04 & 3.61$\pm$0.04 & 3.88$\pm$0.05  \\ 
        \midrule
            DiEmo-TTS & \textbf{3.77$\pm$0.05} & \textbf{3.78$\pm$0.04} & 3.70$\pm$0.04 & \textbf{3.80$\pm$0.04} & \textbf{3.69$\pm$0.05} & \textbf{3.72$\pm$0.05} & \textbf{3.78$\pm$0.04} & \textbf{3.69$\pm$0.05} & \textbf{3.95$\pm$0.04} & \textbf{3.83$\pm$0.04} & \textbf{3.81$\pm$0.05} & \textbf{4.03$\pm$0.05} \\ 
        \bottomrule
    \end{tabular}
      }\vspace{-0.3cm}
\end{table*}

\subsection{Style-adaptive condition modules}
In cross-speaker emotion transfer, a single model synthesizes speech while accounting for both speaker and emotion conditions. Beyond naturalness, maintaining stylistic consistency is crucial for high-quality synthesis. Inspired by \cite{kong2023vits2}, we design a dual conditioning transformer block (DCT) to fuse the multiple factors into the intermediate feature of the encoder and decoder. We apply weight-sharing multi-head attention conditioning for each type of style information on the transformer block. The attention outputs are fused by a multi-layer perceptron, which effectively integrates multiple speech features.

\subsection{TTS model}
We retain the original architecture and objective function of FastSpeech 2 \cite{ren2020fastspeech}, except for the inclusion of an emotion embedding $E_{emo}$ to provide a speaker-disentangled emotional state and the integration of the objective function $L_{\text{DINO}}$. $E_{emo}$ is fixed as one of the short crops of the reference mel-spectrogram corresponding to the text and uses the output of the student encoder. Additionally, the speaker label is mapped into an embedding $E_{spk}$ to represent different speaker characteristics. During inference, we input the desired emotional reference mel-spectrogram into the student encoder to extract the emotion embedding, which is then used for speech synthesis.

\section{Experiments and results}

\subsection{Experimental setup}
We use the emotional speech dataset (ESD) \cite{zhou2022emotional}, which consists of about 350 parallel utterances spoken by 10 English speakers with five emotional states. Following the prescribed data partitioning criteria, we trained ten speakers in total. However, for the cross-speaker emotion transfer scenario, only two target speakers for inference, a male (``0013'') and a female (``0019''), were trained exclusively on neutral utterances.
Moreover, we used the MSP-Podcast corpus dataset \cite{lotfian2017building}, which contains emotional speech data labeled with emotion categories and emotional attribute values. 
We employ the AdamW optimizer with hyperparameters $\beta_{1}=0.9$ and $\beta_{2}=0.98$, and configure the learning rate for the TTS system training at $5\times 10^{-4}$. The training process of the TTS module was conducted over approximately 24 hours on a single NVIDIA RTX A6000 GPU. For the audio synthesis in our experiments, we utilize the official implementation of BigVGAN \cite{lee2023bigvgan}, along with its pre-trained model.

\subsection{Implementation details}
For the acoustic model, we follow the FastSpeech 2 \cite{ren2020fastspeech} configuration, setting the FFT block of the phoneme encoder and decoder to 4 layers, 256 hidden size, 1024 filter size, and 9 kernel size. Additionally, DCT is applied to the third FFT block. For the teacher and student encoders, we use the Mel-style encoder proposed in \cite{min2021meta}. Cluster-driven sampling creates a set of five utterances based on the cluster and randomly selects them. Similar to the configuration in \cite{han2023self}, for each utterance, two long crops (3 seconds) and four short crops (2 seconds) are randomly extracted. Additionally, the DINO projection head and hyperparameters are set accordingly. For the emotional attribute prediction \cite{goncalves2024odyssey}, we use the fine-tuned module that integrates components from a pre-trained WavLM \cite{chen2022wavlm}. Furthermore, we used k-means clustering and ECAPA-TDNN \cite{wang22q_interspeech} for emotion clustering, following the configuration of the original paper \cite{ulgen2024revealing}.

\begin{figure}[!t] 
    \centering
\includegraphics[width=1.0\linewidth]{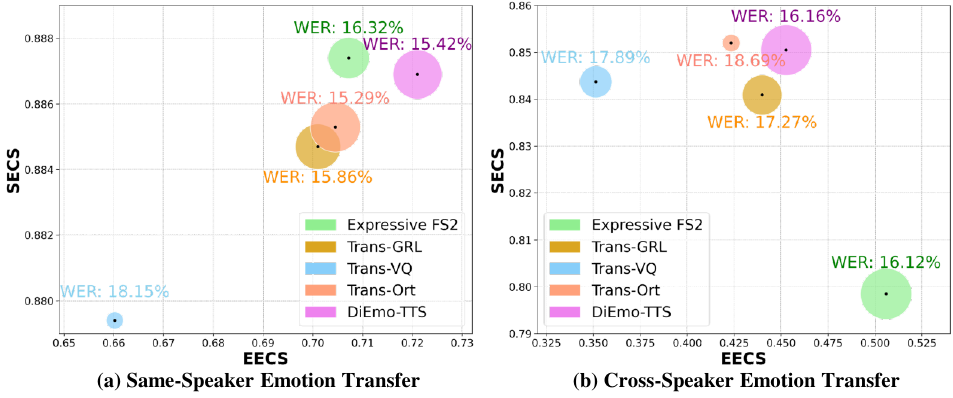}\vspace{-0.3cm}
\caption{Comparision of our proposed method in terms of speaker and emotion similarity}
\label{Figure2}\vspace{-0.3cm}
\end{figure}

\subsection{Performance metrics}
For evaluation metrics, we conducted a MOS evaluation for naturalness (nMOS), speaker similarity (sMOS), and emotion similarity (eMOS) using a five-point scale ranging from 1 to 5. The results are presented with a confidence interval of 95\%. To evaluate linguistic consistency, we calculate the word error rate (WER) and character error rate (CER) by Whisper large model \cite{radford2023robust}. For the speaker similarity measurements, we calculate the speaker embedding cosine similarity (SECS) via WavLM\footnote{https://huggingface.co/microsoft/wavlm-base-sv} between the target and synthesized speech. For emotionally expressive evaluation, we calculate the emotion embedding cosine similarity (EECS) \cite{oh2024durflex} using Emotion2Vec\footnote{https://github.com/ddlBoJack/emotion2vec} \cite{ma2023emotion2vec} between the target emotion and synthesized speech. For subjective evaluation, 20 participants assessed an evaluation set constructed by selecting two random samples per emotion for each speaker. For objective evaluation, the entire evaluation set was used.

\begin{table*}[!ht]
    \centering 
        \caption{Cross speaker emotion transfer results for ablation studies.}
    \label{Table2}\vspace{-0.3cm}
        \resizebox{0.93\textwidth}{!}{
    \begin{tabular}{l|>{\centering\arraybackslash}p{1.1cm}>{\centering\arraybackslash}p{1.1cm}>{\centering\arraybackslash}p{1.1cm}>{\centering\arraybackslash}p{1.1cm}|ccc|cccc}
        \toprule
        \textbf{Method} & \textbf{IP} & \textbf{CS} & \textbf{CDS} & \textbf{DCT} & \textbf{nMOS} ($\uparrow$) & \textbf{sMOS} ($\uparrow$) & \textbf{eMOS} ($\uparrow$) & \textbf{WER} ($\downarrow$) & \textbf{CER} ($\downarrow$) & \textbf{SECS} ($\uparrow$) & \textbf{EECS} ($\uparrow$) \\
        \midrule
            DiEmo-TTS & \cmark & \cmark & \cmark & \cmark & \textbf{4.23$\pm$0.04} & 3.96$\pm$0.03 & \textbf{4.07$\pm$0.04} & \textbf{16.16} & \textbf{5.60} & 0.8505 & \textbf{0.4527} \\ 
        \midrule
            \multirow{4}{10em}{Ablation Study} & \cmark & \cmark & \cmark & \xmark $\cellcolor{gray!25}$ & 4.16$\pm$0.05 & \textbf{4.02$\pm$0.04} & 3.89$\pm$0.05 & 16.63 & 6.15 & \textbf{0.8626} & 0.4056 \\
            & \cmark & \cmark & \xmark $\cellcolor{gray!25}$ & \xmark $\cellcolor{gray!25}$ & 4.18$\pm$0.05 & 3.84$\pm$0.03 & 4.01$\pm$0.04 & 16.72 & 6.05 & 0.8261 & 0.4461 \\
            & \cmark & \xmark $\cellcolor{gray!25}$ & \xmark $\cellcolor{gray!25}$ & \xmark $\cellcolor{gray!25}$ & 4.20$\pm$0.04 & 3.83$\pm$0.04 & 4.00$\pm$0.05 & 17.47 & 6.46 & 0.8246 & 0.4478 \\
            & \xmark $\cellcolor{gray!25}$ & \xmark $\cellcolor{gray!25}$ & \xmark $\cellcolor{gray!25}$ & \xmark $\cellcolor{gray!25}$ & 4.12$\pm$0.03 & 3.81$\pm$0.05 & 3.95$\pm$0.04 & 17.20 & 6.66 & 0.8217 & 0.4357 \\
        \bottomrule
    \end{tabular}
      }\vspace{-0.3cm}
\end{table*}

\subsection{Model performance}
To ensure a fair comparison between the proposed method and existing cross-speaker emotion transfer approaches, we trained the models using the same dataset and configurations as FastSpeech 2 \cite{ren2020fastspeech}, including the style-adaptive condition modules. We replaced the emotion disentangling and transfer method with three competing cross-speaker emotion transfer methods: 1) \textbf{Trans-GRL} \cite{jo2023cross, oh2024durflex}, which employs adversarial speaker training using the GRL \cite{ganin2016domain}; 2) \textbf{Trans-VQ} \cite{zhang2023iemotts, huang2022generspeech}, which implements a bottleneck layer via a modified VQ layer \cite{van2017neural}; and 3) \textbf{Trans-Ort} \cite{li2022cross1}, which constrains the emotion embedding to be speaker-irrelevant using an orthogonal loss \cite{ranasinghe2021orthogonal} based on the GRL. Additionally, to analyze the role of the disentangling method, we trained a version of the proposed method without the disentangling method, referred to as \textbf{Expressive FS2}.

As shown in Table \ref{Table1}, our model achieves the best performance in preserving the target identity in the synthesized speech while also demonstrating the best performance in transferring all emotions. Furthermore,  our model achieves the best performance in objective metrics for both cross-speaker emotion transfer and same-speaker emotion transfer, as shown in Figure \ref{Figure2}. Specifically, in terms of nMOS and WER averaged over all emotion styles, our proposed method outperforms Trans-GRL, Trans-VQ, and Trans-Ort by achieving higher nMOS and lower WER. These results demonstrate that the emotion representations disentangled by self-supervised distillation effectively minimize other information losses while achieving a balance between preserving the target speaker identity and transferring the source emotion.

\begin{figure}[!t] 
    \centering
\includegraphics[width=1.00\linewidth]{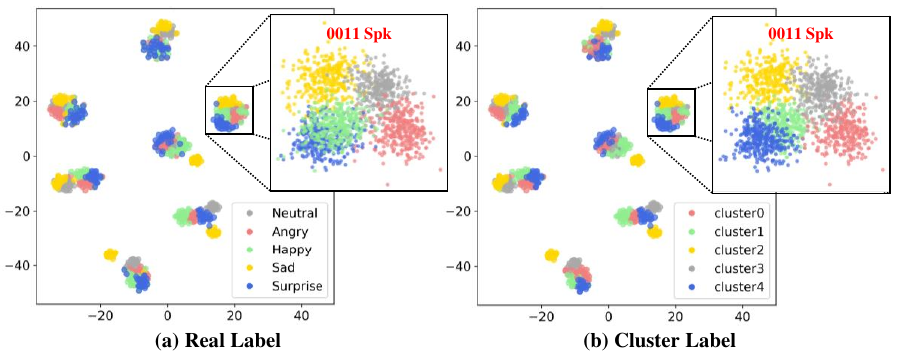}\vspace{-0.3cm}
\caption{t-SNE of speaker embeddings in ESD dataset.}
\label{Figure3}\vspace{-0.2cm}
\end{figure}

\subsection{Analysis of emotion clustering and matching}
\begin{figure}[!t] 
    \centering
\includegraphics[width=0.90\linewidth]{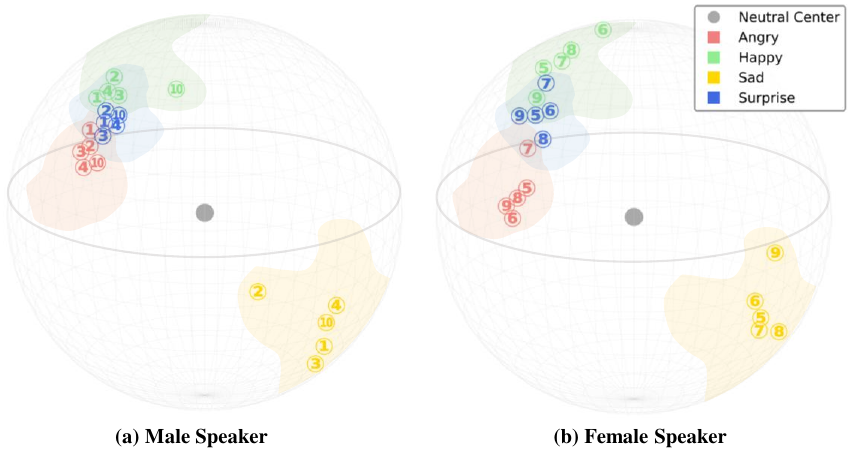}\vspace{-0.3cm}
\caption{Visualization of the relative positional information of emotional clusters, represented by azimuth and elevation angles, for each speaker in spherical coordinates. The numbers in the plot represent individual speakers.}
\label{Figure4}\vspace{-0.3cm}
\end{figure}

To analyze the effectiveness of emotion clustering and matching via emotional attribute prediction, we visualized the ESD dataset \cite{zhou2022emotional} using the pre-trained model \cite{goncalves2024odyssey}. Figure \ref{Figure3} illustrates the distribution of the embeddings, where clusters separate clearly and align well with actual labels. These results indicate that speaker embeddings naturally form distinct groups corresponding to different emotional states due to unique vocal characteristics associated with each emotion \cite{ulgen2024revealing}. The emotional attribute-based approach maintains high matching accuracy, demonstrating its robustness across different domains. Furthermore, as depicted in Figure \ref{Figure4}, the relative positional information of clusters in spherical coordinates remains consistently grouped by emotion across speakers, reinforcing the stability of emotion-based clustering in this approach.

\subsection{Ablation study}
This paper introduces an emotion disentangling method as a key extension to the traditional DINO method. To evaluate its effectiveness, we conduct an ablation study. As shown in Table \ref{Table2}, our model significantly improves performance, explained as follows: 1) ``w/o DCT'' removes the dual conditioning transformer block, replacing it with concatenation conditioning. Unlike simple style transfer, integrating multiple speech features better balances emotion and speaker characteristics, enhancing speech quality. 2) ``w/o CDS'' and ``w/o CS'' denote the removal of cluster-driven sampling, where random cropping is applied to a single sampled utterance, and the exclusion of cosine similarity loss, respectively. The removal of CDS and CS causes the emotion encoder to focus solely on emotional information rather than achieving proper disentangling, resulting in poorer overall performance and higher emotion-related metrics. 3) ``w/o IP'' replaces formant-based information perturbation with noise augmentation using the MUSAN \cite{snyder2015musan} and RIR \cite{ko2017study} datasets. While background noise or reverberation adds distortion, formant perturbation preserves emotion while perturbing speaker identity, improving disentanglement in cross-speaker emotion transfer. These results confirm that the proposed method effectively disentangles emotions and enables the TTS model to transfer multiple emotional and speaker styles harmoniously.

\section{Conclusion}
We present DiEmo-TTS, a system for cross-speaker emotion transfer that employs a novel self-supervised distillation method to minimize emotional information loss while preserving speaker identity. By leveraging a self-supervised distillation method with cluster-based disentangling, our approach effectively minimizes emotional information loss without relying on explicit emotion labels. Additionally, our style-adaptive conditioning enhances expressiveness and stability, producing natural and balanced speech. Experimental results demonstrate that the proposed method represents a new state-of-the-art in learning speaker-irrelevant emotion embeddings, excelling in expressiveness, naturalness, and speaker identity preservation. Despite its advancements, the proposed method does contain the limitation of relying on emotional dimension pseudo-labels, which originate from a predictor trained on labeled data. Future work will explore unsupervised methods to reduce dependence on pre-trained predictors and enhance the model's ability to handle unseen speakers, enabling more robust cross-speaker emotion transfer. These advancements will drive more robust and scalable cross-speaker emotion transfer, enabling expressive and natural speech synthesis across diverse datasets. 

\section{Acknowledgements}
We’d like to thank Ji-Sang Hwang for helpful contributions to our work. 
This work was partly supported by the Institute of Information \& Communications Technology Planning \& Evaluation (IITP) grant funded by the Korea government (MSIT) (Artificial Intelligence Graduate School Program (Korea University) (No. RS-2019-II190079), Artificial Intelligence Innovation Hub (No. RS-2021-II212068), AI Technology for Interactive Communication of Language Impaired Individuals (No. RS-2024-00336673), and Artificial Intelligence Star Fellowship Support Program to Nurture the Best Talents (IITP-2025-RS-2025-02304828)).

\bibliographystyle{IEEEtran}
\bibliography{refs}

\end{document}